\documentclass[12pt]{article}
\usepackage{amscd}
\usepackage{verbatim}
\usepackage{amssymb}
\usepackage{amsmath}
\usepackage{tabularx}
\usepackage{hyperref}

\newcommand{\be}{\begin{equation}}
\newcommand{\ee}{\end{equation}}
\newcommand{\ben}{\begin{eqnarray}}
\newcommand{\een}{\end{eqnarray}}
\newcommand{\ba}{\begin{eqnarray}}
\newcommand{\ea}{\end{eqnarray}}

\newcommand{\bi}{\begin{itemize}}
\newcommand{\ei}{\end{itemize}}

\begin{document}
\begin{center}
\vspace{24pt} { \large \bf Some results in AdS/BCFT } \\
\vspace{30pt}
\vspace{30pt}
\vspace{30pt}
{\bf Mohd Ali\footnote{mohd.ali@students.iiserpune.ac.in}}, {\bf Vardarajan
Suneeta\footnote{suneeta@iiserpune.ac.in}}\\
\vspace{24pt} 
{\em  The Indian Institute of Science Education and Research (IISER),\\
Pune, India - 411008.}
\end{center}
\date{\today}
\bigskip
\begin{center}
{\bf Abstract}
\end{center}
In this paper, we study the $AdS/BCFT$ construction in $AdS_3$ and the BTZ black hole spacetime. We find a new solution to the equation for the End of the World (EoW) brane. The induced metric on this solution is that of $dS_2$ with an isometry group identical to that of $AdS_2$ and the symmetry group of the corresponding BCFT. It also leads to a corresponding EoW brane in the non-rotating BTZ background which is consistent with the periodic identifications of the non-rotating BTZ black hole. The new solution generalizes easily to higher dimensions.


\newpage
\section{\label{sec:level1}Introduction}
AdS/BCFT is a holographic correspondence between anti-de Sitter spacetime in $d+1$ dimensions, $AdS_{d+1}$ and a conformal field theory in $d$ dimensions, $CFT_d$ with a boundary such that a part of conformal symmetry is preserved at the boundary of the CFT \cite{HDOB,ABCFT,HMM}. It is a generalization of the AdS/CFT correspondence because it provides us a way to construct the gravity dual even if the manifold on which the CFT lives has boundaries. In the AdS/BCFT construction, the gravity dual is equipped with an extra boundary $Q$ in addition to the asymptotically AdS boundary $M$, such that the boundary of $Q$ coincides with that of $M$. $Q$ is an End of the World (EoW) brane. Neumann boundary conditions are placed on the $Q$ surface and Dirichlet boundary conditions are placed on $M$. In AdS/BCFT, the gravity dual is obtained by finding the EoW brane which is a solution to Neumann boundary conditions, with induced metric having an isometry group consistent with the symmetry group of the BCFT. $AdS_3 $ and BTZ black holes are the solutions of interest to Einstein's equation in three dimensions in which we want to study EoW branes.  The $Q$ surface for $AdS_3$ and non-compact BTZ spacetime have been constructed in \cite{HDOB,ABCFT}, but the $Q$ brane solution obtained in the BTZ coordinates is not periodic in the angular coordinate of the BTZ spacetime, so it does not go to a $Q$ brane in the BTZ spacetime with global identifications. More general constructions involving two branes in the BTZ background can be found in \cite{geng}.
\newline In this article, we have discussed AdS/BCFT in three dimensions. We have presented a brief introduction to the AdS/BCFT construction in section II. In section III, we have mentioned all the possible solutions to the Neumann boundary condition in $AdS_3$. Many of these have already been derived in \cite{ABCFT,HMM}.  We discuss one new solution that we have found. The induced metric on this surface is that of $dS_2$, which has the same isometry group as $AdS_2$ (this being a feature of two dimensions) and the same symmetry group as the BCFT. In section IV, we have discussed the AdS/BCFT construction for the non-rotating BTZ spacetime. The new surface obtained in $AdS_3$ can be written in BTZ coordinates and we see that it goes over into the EoW brane in the non-rotating BTZ black hole consistent with the periodic identifications defining the black hole. The new surface can easily  be generalised to higher dimensional $AdS$ spacetime. We also briefly discuss the rotating BTZ black hole, for which there is no $Q$ brane solution periodic in the angular coordinate.

\section{AdS/BCFT Construction }
The gravity dual of the $d$ dimensional CFT that lives on manifold $M$ can be obtained by extending $M$ to a $d+1 $ dimensional manifold $N$ in a way that $\partial N=M$ $\cup$ $Q$, $\partial M=\partial Q$ and $Q$ should be homologous to $M$. $Q$ is an EoW brane. The fact that we have a CFT on $M$ requires that $N$ should be a part of $AdS_{d+1}$. In this construction, the Dirichlet boundary condition is imposed on $M$ and Neumann boundary condition on $Q$. We can write the action for the gravity sector as follows:
\begin{equation}\label{1}
    I=\frac{1}{16\pi G_N}\int_N \sqrt{-g}(R-2\Lambda+L_{M})+\frac{1}{8\pi G_N}\int_Q \sqrt{-h}(K+L^Q_m)
\end{equation}
where $\Lambda$ is the cosmological constant on $N$, $L_{M}$ and $L^Q_{M}$ are the matter Lagrangian in the bulk and on the $Q$ surface respectively. $g_{\mu\nu}$  is a metric on $N$  and $h_{ab}$ is an induced metric on $Q$. Let $x^\mu$ and $y^a$ be the coordinates in the bulk and on the $Q$ surface respectively. $K$ is the trace of extrinsic curvature of the $Q$ surface, which is defined as
\begin{align}\label{2}
    K_{ab}&=e^\alpha_a e^\beta_b\nabla_{\alpha}n_{\beta} & K=h^{ab}K_{ab}.
\end{align}
 Here, $e^{\alpha}_a=\frac{\partial x^\alpha}{\partial y^a}$ and $n$ is the normal vector to the $Q$ surface. After varying the above action with Dirichlet boundary conditions on $M$,
\begin{equation}\label{3}
    \delta I= \frac{1}{16\pi G_N}\int_Q \sqrt{-h}(K_{ab}-Kh_{ab}-T^Q_{ab})\delta h^{ab} .
\end{equation}
The Neumann boundary condition on $Q$ implies:
\begin{equation}\label{4}
    K_{ab}-Kh_{ab}-T^Q_{ab}=0
\end{equation}
Recall that for a $d$ dimensional CFT without boundary, $CFT_d$, the conformal symmetry group is $SO(2,d)$. In AdS/BCFT, the BCFT is a $d$ dimensional CFT with a $d-1$ dimensional boundary, such that the boundary CFT has $SO(2,d-1)$ symmetry.\\
Now, we want to consider a simple class\footnote{Q surface equation for the general stress-energy tensor discussed in later section.} of $Q$ surfaces where $L^Q_M$ is a constant. This is the stress-energy tensor that has been considered in previous work.  This gives
\begin{equation}\label{5}
    K_{ab}=(K-T)h_{ab}
\end{equation}
where $d.T$ is the trace of $T_{ab}^Q$ and $T$ can be thought of as brane tension. Taking the trace of the above equation we get
\begin{equation}\label{6}
    K=\frac{d}{d-1}T
\end{equation}
We can calculate the extrinsic curvature in the bulk coordinates by writing the induced metric of the $d$ dimensional $Q$ surface $h_{ab}$ in terms of a $2$-tensor $ h_{\mu \nu}$ (projection tensor) in the $d+1$ dimensional bulk coordinates as
\begin{align}\label{7}
    h_{\mu\nu}&=g_{\mu\nu}-\epsilon n_\mu n_\nu &             h_{ab}&=e^\mu_a e^\nu_b h_{\mu \nu}.
\end{align}
Here $n^\mu$ is normal to the surface, $n^\mu n_\mu=\epsilon$ where $\epsilon$ is either $1$ or $-1$ depending on whether the normal is spacelike or timelike. $\mu, \nu =(0,1,2...,d)$ are indices denoting the coordinate in the bulk spacetime. Then, the extrinsic curvature  is defined as
\begin{equation}\label{8}
    K_{\mu\nu}=h^\alpha_\mu h^\beta_\nu\nabla_\alpha n_\beta .
\end{equation}
The equation for Neumann boundary condition in the bulk spacetime coordinates becomes
\begin{equation}\label{9}
    K_{\mu \nu}=(K-T)h_{\mu\nu}
\end{equation}
The above expression is equivalent to (\ref{5}) after projecting it on to the surface $Q$. $K_{\mu\nu}n^\nu=h_{\mu\nu}n^\nu=0$ follows from the definition of extrinsic curvature and the induced metric.

\section{Q surface in Poincare$'$ AdS$_3$}
We can classify the solutions to (\ref{9})  in terms of the number of non-zero components of the normal and the value of $(Tl)^2$. The metric of Poincare $AdS_3$ is
\begin{equation}\label{10}
     ds^2=\frac{l^2}{z^2}(dz^2+dx_1^2-dx_0^2)
\end{equation}
The equation of the $Q$ surface can be denoted as the level set $F=0$, where $F$ depends on the bulk coordinates. The normal to the surface is $n_{\mu} = \epsilon \frac{\partial_{\mu} F}{\mathcal{N}}$ where $\mathcal{N}^2 = g^{\mu \nu} \partial_{\mu} F \partial_{\nu} F $ is the normalization factor. Surfaces with $n_z=0$ are not solutions to the equation (\ref{9}) because they do not satisfy $K_{zz}=Th_{zz}$\footnote{ $K_{zz}=\partial_zn_z-\Gamma^z_{zz}n_z=0$ when $n_z=0$ but $h_{zz}=g_{zz}$ which is non zero.}.The surface $z=$constant does satisfy (\ref{9}) with $T=\pm\frac{1}{l}$, but it does not go up to the boundary of $AdS$. In the AdS/BCFT correspondence, the boundary of the $Q $ surface must coincide with the boundary of the CFT. So, we need to solve equation (\ref{9})  only for those surfaces for which $n_z\neq0$ and out of $n_{x_0}$ and $n_{x_1}$ at least one must be non-zero.
\\
Based on this classification, the solution with $n_{x_0}=0$ corresponds to surfaces of the type $F(x_1,z)$, which were obtained by Takayanagi in \cite{HDOB}. These EoW branes are slices in the slicing of $AdS$ spacetime by $AdS$ slices in one lower dimension. The isometry group of these solutions is the same as that of the boundary CFT. There is also a class of EoW branes with none of the normals being zero. We denote these surfaces by $F(x_0,x_1,z)$ and they are discussed in the section $7$ of \cite{ABCFT} by Fujita, Takayanagi and Tonni (FTT) as also by Akal, Kusuki, Shiba, Takayanagi and Wei (AKSTW) in \cite{HMM}.
\\
\\
There is another class of EoW branes with $n_{x_1}=0$ which we study next. We derive these solutions in the BTZ coordinates in the Appendix (and they go over to $Q$ branes in $AdS_3$ as well).\\
\textbf{ F(z,$\mathbf{x_0}$) surfaces:}\\
We can obtain the following solutions to equation (\ref{9}),\\
when \textbf{ $\epsilon =1$}\\
\begin{equation}\label{20}
    x_0=\frac{Tl}{\sqrt{(Tl)^2-1}}z
\end{equation}
where $(Tl)^2>1$ and now we can parametrize the above solution in the following way,
\begin{align}\label{21}
    z&=y\sqrt{(Tl)^2-1} & x_0&=y Tl .
\end{align}
With this parametrization, the induced metric is
\begin{equation}\label{22}
    ds^2=\frac{l^2}{((Tl)^2-1)}\frac{1}{y^2}\Big(-dy^2+dx_1^2\Big).
\end{equation}
The induced metric is that of two dimensional de Sitter space $dS_2$, which has isometry group identical to that of $AdS_2$. The solution in (\ref{20}) is a new solution with isometry group same as the symmetry group of the BCFT. This surface is a slice in the slicing of $AdS_3$ with $dS_2$ slices. We can easily generalize this result to higher dimensions. It is always possible to foliate $AdS_{d+1}$ with $dS_d$ slicing. In $dS$ slicing coordinates $AdS_{d+1}$ metric with $AdS$ radius $R$ can be written as
\begin{equation}\label{22*}
 ds^2= d\rho^2+ \sinh^2({\frac{\rho}{R}})ds^2_{dS_{d}}.
\end{equation}
The EoW brane is given by  $\rho=\rho*$ with  extrinsic curvature $K_{ab}=\frac{1}{R}(\coth\frac{\rho}{R})h_{ab}$, where $h_{ab}$ is the induced metric on the EoW brane and $\rho*$ is defined by
 \begin{equation}
 T=\frac{d-1}{R}\coth \frac{\rho*}{R}.
 \end{equation}
When \textbf{$\epsilon=-1$}, we obtain
\begin{equation}\label{23}
    x_0=\frac{Tl}{\sqrt{(Tl)^2+1}}z
\end{equation}
here $Tl$ can take any value. Using similar parametrization as above,
\begin{align}\label{24}
     x_0&=y Tl & z=\sqrt{(Tl)^2+1}y
\end{align}
we can get the induced metric on this surface as
\begin{equation}\label{25}
    ds^2=\frac{l^2}{((Tl)^2+1)}\frac{1}{y^2}\Big(dy^2+dx_1^2\Big)
\end{equation}

As we saw, the induced metric on the brane (\ref{22}) is that of $dS_2$. Constructions where the brane is a Friedmann-Lemaitre-Robertson-Walker (FLRW) metric can be found in \cite{swingle} where it is proposed to study cosmology on the brane using AdS/BCFT.
\section{Q surfaces in the non-rotating BTZ Black hole}
The BTZ black hole is a three dimensional solution to Einstein's gravity with negative cosmological constant. It is obtained by doing discrete identifications in $AdS_3$ by a discrete subgroup of the isometry group\cite{TBHTD,GBH}. The metric of the non-rotating BTZ spacetime is
\begin{equation}\label{26}
    ds^2=-\frac{r^2-r^2_0}{l^2}dt^2+\frac{l^2}{r^2-r^2_0}dr^2+r^2dx^2
\end{equation}
where $x$ is periodic with period $2\pi$. Following the same procedure as in the last section, we can classify the solutions to (\ref{9}) depending on the non-zero components of the normal and value of $(Tl)^2$. Every solution of (\ref{9}) in the BTZ spacetime with global identifications will also be a solution in $AdS_3$ (by unwrapping the periodic coordinate). The converse is not true in general, since we need the solution to the Neumann boundary condition in the BTZ black hole to be periodic in the $x$ coordinate or $x$-independent. If this were not true, then an explicit identification of $x$ with $x+ 2n\pi$ implies we have to identify two points on the surface with different $r$ values as well. Then $Q$ may not be homologous to $M$.\\
\textbf{1. F(r,x) Surface}\\
For the above surface the $n_t$ component of the normal vanishes and solving (\ref{9}) for $F(r,x)$ in BTZ spacetime gives solutions for different values of $(Tl)^2$, as shown by AKSTW in \cite{HMM}.   \\
For $(Tl)^2<1$ we will get,
\begin{equation}\label{27}
    r=\frac{Tlr_0}{(1-(Tl)^2)^\frac{1}{2}}\frac{1}{\sinh(\frac{r_0x}{l})}
\end{equation}
 One can similarly obtain solutions for $(Tl)^2=1$ and $(Tl)^2>1$ as obtained in \cite{HMM}. AKSTW have considered non-compact BTZ, i.e the BTZ metric where $x$ is a non-compact direction. In our case, the $x$ in the BTZ metric is periodic with a period $2\pi$. These solutions will be EoW branes in the BTZ black hole spacetime only when consistent with the periodicity of the coordinate $x$ where $x$ is identified with $x + 2n\pi$, $n$ an integer. As can be seen, (\ref{27}) is not periodic in $x$, neither are the solutions for $(Tl)^2=1$ and $(Tl)^2>1$.  \\Also, note that in the equation (\ref{27}) we can write $x$ in terms of $r$, and then we can easily show that $x$ can be greater than $2\pi$ even if we restrict ourselves outside the horizon.
 \footnote{In (\ref{27}) with $r>r_0$, $x<\frac{l}{r_0}\sinh^{-1}{\frac{Tl}{\sqrt{1-(Tl)^2}} }
$ where $-\infty<\frac{Tl}{\sqrt{1-(Tl)^2}}<\infty$. Therefore, $x$ can be greater than $2\pi$. }

In three-dimensional gravity with a negative cosmological constant, every solution to Einstein's equation is $AdS_3$ quotiented by some subgroup of the isometry group. So any solution in three dimensions locally looks like $AdS_3$. Under the following transformation, we can write the coordinate transformation from Poincare-$AdS_3$ coordinates to the non-rotating BTZ coordinates:
\begin{align*}
    x_0-x_1&=-e^{\frac{r_0(lx-t)}{l^2}}\sqrt{1-\frac{r^2_0}{r^2}}\\
\end{align*}

\begin{align}\label{30}
     x_0+x_1&=e^{\frac{r_0(lx+t)}{l^2}}\sqrt{1-\frac{r^2_0}{r^2}}
\end{align}
\begin{align*}
    z&=\frac{r_0}{r}e^{\frac{r_0x}{l}}\\
\end{align*}

Using the above coordinate transformation the $F(r,x)$ class of solutions can be mapped to  the $F(x_1,x_0,z)$  class of solution in $AdS_3$. Note that after the coordinate transformations between the Poincare-$AdS_3$ coordinates and the BTZ coordinates, we need to make global identifications such that $x$ is periodic with period $2\pi$ to obtain the surface in the BTZ spacetime.  \\
\textbf{2. F(r,t) Surface}\\
We can obtain the $F(r,t)$ surface either by following the same steps as in the last section or we can use (\ref{30}) to get the surface $F(r,t)$ from the known surface in $AdS_3$. Following any of the above procedures, we will get,\\
\\
when $(Tl)^2<1$ and $\epsilon=1$
\begin{equation}\label{31}
    \cosh{(\frac{r_0t}{l^2})}\sqrt{r^2-r_0^2}=\frac{Tlr_0}{\sqrt{1-(Tl)^2}}
\end{equation}
when $(Tl)^2>1$ and $\epsilon=1$
\begin{equation}\label{32}
    \sinh{(\frac{r_0t}{l^2})}\sqrt{r^2-r_0^2}=\frac{Tlr_0}{\sqrt{(Tl)^2-1}}
\end{equation}
when $\epsilon=-1$
\begin{equation}\label{33}
     \sinh{(\frac{r_0t}{l^2})}\sqrt{r^2-r_0^2}=\frac{Tlr_0}{\sqrt{(Tl)^2+1}}.
\end{equation}
Using the transformations in (\ref{30}), we can show that solution (\ref{31}) can be mapped to the $F(x_1,z)$ class of surfaces in $AdS_3$ found by Takayanagi, (\ref{32}) to (\ref{20}), and (\ref{33}) to (\ref{23}). The surface in equation (\ref{32}) is the transformed new solution we found in $AdS_3$ with a $dS_2$ metric induced on it. All the above solutions are independent of the angular coordinate $x$ and are consistent with the global identifications. The explicit calculation for the new surface is given in appendix \ref{A2}. In a similar manner, we can obtain an $F(r,t,x)$ class of solutions (with all the normal components non-zero) from corresponding solutions in $AdS_3$ and this class will not be periodic in $x$. \\
\\
\textbf{3. Rotating BTZ Black holes}\\
Similar to the case of BTZ black holes, one can use coordinate transformations that take the rotating BTZ black hole metric to the $AdS_3$ metric to obtain EoW branes in the noncompact rotating BTZ spacetime from the branes in $AdS_3$. It can be checked that none of the solutions is periodic in $x$ or independent of $x$. So none of these solutions are $Q $ branes in the rotating BTZ spacetime with global identifications. More general stress-energy tensors have been considered in the literature and the corresponding branes in $AdS$ have been found \cite{magan}, \cite{flory}. It would be interesting to see what sort of stress energy tensor would yield a brane in the rotating BTZ background, consistent with periodicity of $x$.

\section{Conclusion and Discussion}
In this paper, we have explored the construction of gravity dual to BCFT via AdS/BCFT correspondence in three dimensions. We have obtained a new $Q$ brane solution to the Neumann boundary condition in $AdS_3$, with SO(2,1) isometry group which is the same as the symmetry group of the BCFT. The new solution can also be generalized to a EoW brane in higher dimensional $AdS$ spacetime. In three-dimensional gravity with a negative cosmological constant, every solution to the Einstein equation is $AdS_3$ up to a discrete identification by some subgroup of the isometry group. So every solution to the Neumann boundary condition in $AdS_3$ will also be a solution in BTZ spacetime provided it is periodic in the angular coordinate of the BTZ spacetime. We have shown that there exists a $Q$ surface solution to the Neumann boundary condition in the non-rotating BTZ spacetime consistent with the identification. All the other potential $Q$ surfaces for the BTZ spacetime are not periodic in the angular coordinate $x$.  If we make any explicit identifications of points in the $Q$ brane, we have to identify points with different values of $r$ and then the surface may not be homologous to $M$. In the case of the rotating BTZ black hole, there is no solution periodic in $x$ or independent of $x$ with the choice of stress energy tensor proportional to the metric. So there may be no $Q$ brane in the rotating BTZ black hole consistent with the periodic identifications in that case.
\section{Appendix}

\subsection{Calculation of Q surface equation in BTZ Spacetime}\label{A2}
Let us construct the non static solution to equation (\ref{9}) in BTZ spacetime. Let the surface be $f(r,t)=0$, then the unit normal vector $n^\mu$ is defined as
\begin{equation}\label{1A}
    n_\mu= \frac{\epsilon\partial_\mu f}{(\frac{r^2-r^2_0}{l^2}(\partial_rf)^2-\frac{l^2}{r^2-r^2_0}(\partial_tf)^2)^\frac{1}{2}}.
\end{equation}
Here $n$ is normalized as
\begin{equation}\label{2A}
    n^\mu n_\mu=\epsilon.
\end{equation}
\begin{align*}
    \epsilon&= +1\hspace{3mm}spacelike\\
    \epsilon&=-1\hspace{2mm}timelike
\end{align*}
Equation (\ref{9}) in three dimensions can be written as
\begin{equation}\label{3A}
    K_{\mu\nu}=Th_{\mu\nu}
\end{equation}
where
\begin{equation}\label{4A}
    K_{\mu\nu}=h^\alpha_\mu h^\beta_\nu \nabla_\alpha n_\beta .
\end{equation}
\begin{equation}\label{5A}
    h_{\mu\nu}=g_{\mu\nu}-\epsilon n_\mu n_\nu
\end{equation}
where $g_{\mu\nu}$ is the BTZ black hole metric. we can easily show that from the $xx$ component of the equation (\ref{9}),
\begin{equation}\label{6A}
    n_r=\frac{Tl^2r}{r^2-r_0^2}.
\end{equation}
Using the above equation and $rr$ component of the (\ref{9}), we obtain
\begin{equation}\label{7A}
    n_t^2=T^2r^2-\epsilon\frac{r^2-r_0^2}{l^2}.
\end{equation}
The rest of the equations in (\ref{9}) will trivially be satisfied. To determine the equation of the surface, consider
\begin{equation}\label{8A}
    \frac{n_t}{n_r}=\frac{\partial_t f}{\partial_r f}
\end{equation}
for $(Tl)^2>1$. We can define a new coordinate $r'$ as follows:
\begin{equation}\label{9A}
    r'=-\frac{l^2}{r_0}\sinh^{-1}\Big(\frac{Tlr_0}{\sqrt{(Tl)^2-\epsilon}}\frac{1}{\sqrt{r^2-r_0^2}}\Big).
\end{equation}
In $r', t$ coordinates, equation (\ref{9}) will become
\begin{equation}\label{9B}
    (\partial_{r'}-\partial_t)f=0
\end{equation}
which has the general solution
\begin{equation}\label{10A}
    f=f(r'+t).
\end{equation}
Since  the surface equation is $f(r'+t)=0$, $r'+t=c$ is the solution for any $c$ which is the root of the equation $f(r'+t)=0$. The equation of the surface in $(r,t)$ coordinates is
\begin{equation}\label{11A}
    \frac{Tlr_0}{\sqrt{(Tl)^2-\epsilon}}=\sqrt{r^2-r_0^2} \sinh(\frac{(t+c)r_0}{l^2}).
\end{equation}
The rest of the solutions can also be obtained in similar manner for different $(Tl)^2$ .

\section{Acknowledgements}
MA thanks Tadashi Takayanagi for a useful clarification on his previous work. We thank Mario Flory for bringing his work on general stress energy tensors in AdS/BCFT to our attention.
MA also acknowledges the Council of Scientific and Industrial Research (CSIR),Government of India for financial assistance.

\end{document}